\documentclass[8pt,a4paper]{article}

\usepackage[utf8]{inputenc}
\usepackage[T1]{fontenc}
\usepackage{lmodern}
\usepackage{amsmath,amssymb}
\usepackage{graphicx}
\usepackage{color}
\usepackage[colorlinks,linkcolor=red,anchorcolor=green,citecolor=blue]{hyperref}
\usepackage{cite}
\usepackage{authblk}
\usepackage{booktabs}
\usepackage{array}
\usepackage{multirow}
\usepackage{geometry}
\usepackage{lineno}
\usepackage{bm}
\usepackage{ulem}

\usepackage{setspace}
\title{\textbf{From Hits to Tracks: A BERT-based Tracking Model for Track Reconstruction in Drift Chambers}}

\author[1,2]{Yipu Liao}
\author[1,2,*]{Yao Zhang}
\author[1]{Jingde Chen}
\author[3,*]{Shimiao Jiang}
\author[1,2]{Liyan Qian}
\author[1,2]{Zhaoke Zhang}
\author[1,2]{Ye Yuan}
\author[1,2]{Ke Li}
\author[1,2]{Changzheng Yuan}

\affil[1]{Institute of High Energy Physics, Chinese Academy of Sciences, Beijing}
\affil[2]{University of Chinese Academy of Sciences, Beijing}
\affil[3]{China Academy of Space Technology, Beijing}
\affil[*]{Corresponding authors: zhangyao@ihep.ac.cn,shimiao\_jiang@163.com}

\date{\today}

\begin{document}
% \linenumbers

\maketitle

\begin{abstract}
Track reconstruction in drift chambers is essential for momentum measurement and particle identification at electron-positron colliders. While Transformer architectures have transformed many sequence-processing domains, their application to tracking in high energy physics is still being explored. We present a model that combines a BERT encoder with a Transformer decoder to perform hit-to-track association through autoregressive sorting. The model is evaluated on the DCTracks open dataset that provides realistic drift chamber simulations with varying particle types, momenta, track multiplicities, and noise conditions. Across single-track, two-track, and multi-track samples, the model achieves high hit and track efficiencies while keeping the rates of clones and fakes very low. It also works well in the reconstruction of displaced vertices. These results show BERT-based sequence-to-sequence models as a promising approach for track reconstruction in low-background, precision-oriented experiments.
\end{abstract}

%=============================================================================
\section{Introduction}
%=============================================================================

Precision tests of the Standard Model and searches for New Physics depend on accurate reconstruction of charged particle trajectories. Track reconstruction, assigning detector hits to individual particles and estimating their kinematic parameters, is a core and computationally demanding step in experimental data analysis at electron-positron colliders. Traditional approaches at Drift Chambers (DC) use combinatorial pattern recognition and Kalman filter-based fitting~\cite{Fruhwirth:1987fm,Wang:2009kx,Yao:2007kx,Liu:2008kx,Ma:2013ipa,Zhang:2018okf}. Traditional algorithms require extensive domain expertise and careful hyperparameter tuning, leading to long development cycles. Moreover, they are often strongly dependent on the detector geometry, requiring redevelopment for different experiments or software frameworks, even when using the same detector type.

In recent years, deep learning has shown strong promise for track reconstruction, especially graph neural networks (GNNs)~\cite{Reuter:2024kja,Correia:2025deq,Qian:2026pbd}. GNNs can learn directly from raw detector data and naturally handle irregular detector geometries, which makes them suitable for end-to-end learning of track parameters such as momentum, direction, and associated hits~\cite{Qasim:2019otl,Kieseler:2020wcq}. However, they usually require explicit graph construction and can be difficult to train~\cite{Reuter:2024kja}. At the same time, Transformer architectures~\cite{Vaswani:2017kx} have achieved great success in natural language processing, computer vision, and more recently in scientific applications. The BERT model~\cite{Devlin:2018mgb} uses bidirectional self-attention to collect global context from an entire input sequence. For particle tracking, this means that BERT can model correlations among all hits in an event, which makes it possible to perform denoising and track finding in a single framework. Ref.~\cite{Melkani:2024yuj} first proposed treating track finding as a sorting problem in the TrackSorter algorithm. TrackSorter uses a Transformer encoder-decoder to reorder space points according to track membership, and it showed strong performance on the TrackML dataset~\cite{Amrouche:2019wmx}, which is based on proton-proton collisions at the HL-LHC. Deep learning methods have also been increasingly explored for particle reconstruction at electron-positron colliders, where drift chambers are widely used for charged particle tracking. Recently, Ref.~\cite{Qian:2026pbd} released a DC-based dataset, which is an open dataset of drift chamber simulations based on the BESIII detector, and makes it possible to apply machine-learning methods to tracking tasks in this energy region.

In this work, we present a BERT-based Transformer model for track reconstruction in drift chambers and extend the sequence sorting approach to drift chamber data\footnote{The previous preliminary work was presented in \href{https://indico.cern.ch/event/1471803/contributions/6967189/}{CHEP 2026}.}. The model is evaluated on the DCTracks dataset~\cite{Qian:2026pbd}. The relevant codes for this work are available
on GitHub at \url{https://github.com/liaoyp0615/TrackBERT}.

%=============================================================================
\section{Related Work}
%=============================================================================

\subsection{Transformer-based Approaches for Tracking}

Transformer architectures have recently drawn attention in particle tracking. Ref.~\cite{Melkani:2024yuj} uses a standard Transformer encoder-decoder and has been demonstrated on the TrackML dataset. It formulates track finding as a sequence-to-sequence sorting problem, where hits are initially ordered by their distance from the collision point, and the model learns to reorder them by track membership, inserting separator tokens between tracks. Ref.~\cite{Dispoto:2025zyz} developed a Transformer classifier for pileup hit rejection in the MEG II experiment~\cite{MEGII:2018kmf}, which employs a drift chamber as the tracking detector. Their results show that attention mechanisms can effectively separate signal from background even at high cell occupancies of 35-50\%. Ref.~\cite{Caron:2024cyo} explores three different Transformer models, a full encoder-decoder, encoder-only with a classifier, and encoder-only with a regressor, to perform track association predictions for collision event hit points. This work uses the REDVID simulation framework~\cite{Odyurt:2023zmt} along with reduced versions of the TrackML dataset to create five datasets ranging from simple to complex scenarios. Ref.~\cite{VanStroud:2024fau} applies a full Transformer architecture in which the encoder processes the input hits and the decoder performs hit filtering and track reconstruction. Evaluated on the TrackML dataset, this approach achieves a tracking efficiency of 97\% and a fake rate of 0.7\% for particles with transverse momentum above 750 MeV.

\subsection{Tracking Datasets}

This work uses the DCTracks dataset~\cite{Qian:2026pbd}, which was created to address the lack of standardized DC datasets for machine-learning-based tracking. Drift chambers at electron-positron colliders generally have lower event multiplicity, cleaner event topologies, and stricter requirements on momentum resolution and tracking efficiency, especially for particles with low momentum. On the other hand, most existing datasets, such as TrackML~\cite{Amrouche:2019wmx} and ColliderML~\cite{Elitez:2025kx}, are designed for hadron-collider environments. These environments feature high pileup and high multiplicity, with about $10^4$ particles and $10^5$ hits per event.

%=============================================================================
\section{Drift Chamber and Dataset}
%=============================================================================

The drift chamber is tasked with measuring the momentum and position of tracks from final-state charged particles and identifying particle species through the ionization energy loss ($dE/dx$) of charged particles in the gas. It is widely adopted in high energy physics experiments, including BESIII~\cite{BESIII:2009fln}, Belle II~\cite{Belle-II:2010dht}, COMET~\cite{COMET:2018auw}, MEG II~\cite{MEGII:2018kmf}, and FCC-ee~\cite{FCC:2025lpp}. In particular, the central tracker of the BESIII detector is a cylindrical Multi-layer Drift Chamber (MDC) operated in a 1.0~T solenoidal magnetic field. The MDC is 2400~mm long, with an inner radius of 59~mm and an outer radius of 800~mm, covering a polar-angle acceptance of $|\cos\theta| < 0.93$. It contains 6,796 drift cells arranged in 43 sense-wire layers and organized into 11 superlayers. The superlayers alternate between axial and stereo wire orientations to provide three-dimensional position measurements. The design single-wire spatial resolution is approximately 130~$\mu$m, while the transverse momentum resolution is $\sigma_{p_T}/p_T \approx 0.5\%$ at 1~GeV/$c$. More details of the MDC structure can be found in Ref.~\cite{BESIII:2009fln}.

Unlike pixel detectors, which provide dense and regularly sampled two-dimensional measurements, DC produces intrinsically sparse hits from discrete sense wires. Moreover, the drift distance provides an indirect spatial measurement, while axial and stereo wires introduce additional geometrical constraints for three-dimensional track reconstruction. Track reconstruction therefore requires associating sparse hits with trajectories while accounting for detector geometry, hit ordering, and background noise. These characteristics motivate a representation that preserves the sequential and geometrical structure of the hits rather than treating them as a conventional image, providing a natural input format for Transformer-based architectures such as BERT. Accordingly, we formulate the DC hits as structured token sequences for track reconstruction and classification.

The DCTracks dataset~\cite{Qian:2026pbd} provides pre-processed simulated data with realistic detector response and noise overlay. It covers the low-background, low-multiplicity environment of BESIII~\cite{BESIII:2009fln}, a $\tau$-charm factory. The dataset is based on realistic {\sc geant4} Monte Carlo simulations~\cite{Deng:2007zzb} with a full BESIII drift chamber response, including beam-induced backgrounds and electronic noise. Standardized evaluation metrics are provided to enable a fair comparison across different methods.

To facilitate downstream fine-tuning of the BERT-based model, we format the training samples as question-answer pairs. Table~\ref{tab:datasets} summarizes the event samples used in this work. The question is the raw hit map, while the answer contains the hits grouped by type. Noise hits are listed first, followed by track candidates separated by \texttt{[TRACK]} tokens. Hits within each track are sorted by their distance along the trajectory. Each hit has eight features: drift distance and its uncertainty, local and global wire indices, wire-center coordinates at the two MDC end faces, and a truth-matched track ID for supervised learning. The input hits are organized using a radar-like sectorized scheme rather than the conventional circular-range format. The hits are divided into azimuthal sectors and ordered by increasing layer index within each sector, providing a consistent sequence for the model to learn track patterns. The data are split into training, validation, and test sets with a ratio of 14:3:3. Samples 1--5 are used to evaluate the model on clean event topologies. Sample 6 (mixed) tests its performance on different event topologies in a single training set, while Sample 7 ($K_S^0$) evaluates displaced-vertex reconstruction, where the $K_S^0$ decays away from the primary vertex and produces tracks that do not point to it.

\begin{table}[htbp]
\centering
\caption{Summary of the event samples used in this work.}
\label{tab:datasets}
\begin{tabular}{lcccc}
    \toprule
    Sample & Type & Momentum (GeV/$c$) & Number of events \\
    \midrule
    S1 & $e^{\pm}$ & $p \in [0.15, 1.5]$ & 160k \\
    S2 & $\pi^{\pm}$ & $p \in [0.15, 1.5]$ & 160k \\
    S3 & $\mu^+\mu^-$ & $p \in [0.15, 1.5]$ & 120k \\
    S4 & $\pi^+\pi^-J/\psi(\to e^+e^-)$ & $p \in [0.15, 1.5]$ & 175k \\
    S5 & $e^{\pm}$ (low $p_T$) & $p \in [0.04, 0.16]$ & 200k \\
    S6 & Mixed (S1--5) & $p \in [0.04, 1.5]$ & 800k \\
    S7 & $K_S^0\to\pi^+\pi^-$ & $p_{K_S^0} \in [0.2, 2.0]$ & 50k \\
\bottomrule
\end{tabular}
\end{table}

%=============================================================================
\section{Methodology}
%=============================================================================

\subsection{Problem Formulation}

Track reconstruction can be viewed as a sequence-to-sequence sorting problem. The goal is to reorder an unordered set of hits into a structured sequence, where hits that belong to the same particle are grouped together. This formulation fits naturally with Transformer architectures~\cite{Vaswani:2017kx}, which have proven highly effective at sequence transduction tasks.

Given $N$ detector hits, each represented by a multi-dimensional feature vector, the model performs two tasks jointly: identifying noise hits and grouping signal hits into track candidates corresponding to individual charged particles. These tasks are combined into a single sequence prediction framework rather than treated as a cascade. The input is a padded hit sequence $\mathcal{H}={h_1,h_2,\ldots,h_N}$, and the target is a deterministic reordered sequence,
\[
\mathcal{T} = \{h_{n_1}, h_{n_2}, \ldots, \texttt{[TRACK]}, h_{t_{1,1}}, h_{t_{1,2}}, \ldots, \texttt{[TRACK]}, h_{t_{2,1}}, \ldots\},
\]
where noise hits are placed first and signal hits are grouped by track. The learned \texttt{[TRACK]} token acts as a track-boundary marker between consecutive track groups. Within each track, hits are ordered according to their path length along the particle trajectory. This deterministic ordering is important for supervised learning because it provides a unique target sequence for each event.

\subsection{Model Architecture}

\begin{figure}[htbp]
    \centering
    \includegraphics[width=0.99\textwidth]{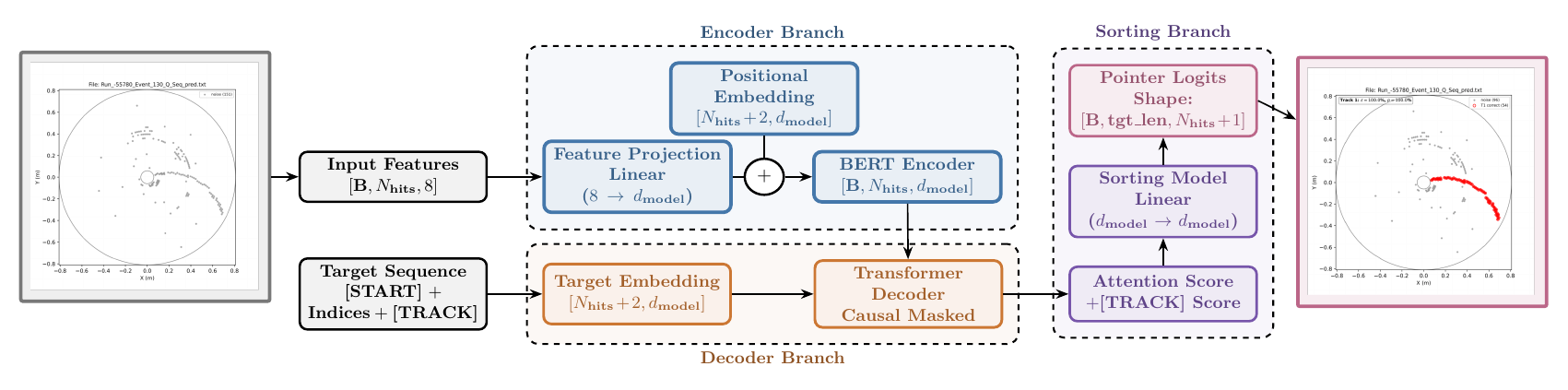}
    \caption{The overview of model architecture. The encoder branch (blue) projects hit features and processes them through a BERT encoder. The decoder branch (orange) takes the target sequence, applies causally masked Transformer decoder layers, and cross-attends to the encoder output. The sorting head (purple) computes pointer logits via attention scoring.}
    \label{fig:architecture}
\end{figure}

The model consists of three components: a BERT-based encoder, a Transformer decoder, and a task-specific sorting head, as shown in Fig.~\ref{fig:architecture}. The encoder provides a global representation of the detector hits, while the decoder establishes a correspondence between the decoder state and the input hits through cross-attention. The sorting head enables reordering of the hit map, thereby facilitating denoising and track finding.

The entire model has $1.6 \times 10^8$ trainable parameters, with the majority coming from the original BERT-based model, which contains $1.1 \times 10^8$ parameters~\cite{Devlin:2018mgb}. Although this parameter count is relatively large, as will be discussed in detail below, we only perform fine-tuning rather than training from scratch; therefore, the number of parameters actually updated is far smaller than this order of magnitude.

Input features with dimension $\mathbb{R}^{B\times N_{\mathrm{max}}\times8}$ are first projected from the eight-dimensional physical feature space to $d_{\mathrm{model}}=768$ by a linear layer. Learnable absolute positional embeddings of size $[N_{\mathrm{max}}+2,d_{\mathrm{model}}]$ are then added to the projected hit features. The resulting sequence is processed by the BERT encoder to obtain contextualized hit representations.

During training, the target sequence is constructed by prepending a learnable \texttt{[START]} token and inserting labeled \texttt{[TRACK]} tokens at track boundaries. An autoregressive Transformer decoder with causal self-attention takes the previously generated target representations together with the encoder output as memory. A pointer scoring head projects the decoder hidden states of shape $\mathbb{R}^{B\times \text{tgt\_len} \times d_{\text{model}} }$ into query vectors and computes scaled dot-product scores against all source hits. Padded source hits are masked by assigning them a score of $-\inf$. In parallel, a learnable track-boundary vector provides an independent score for the \texttt{[TRACK]} token. The hit-selection scores and track-boundary score are concatenated to form the final logits with dimension $\mathbb{R}^{B \times \text{tgt\_len} \times d_{\text{model}} }$. This design allows us to retain the contextual representation capability of BERT while adapting the model to the structured output required by track reconstruction.

\subsection{Training and Inference}

The AdamW optimizer~\cite{Loshchilov:2017} with decoupled weight decay is adopted. We assign a weight decay coefficient $\lambda=0.01$ to most trainable parameters, while excluding biases and LayerNorm weights from regularization with zero weight decay. A linear learning rate scheduler with warm-up is applied: the learning rate rises linearly from zero to the peak value $5 \times 10^{-5}$ over 1,000 steps, then decays linearly until the end of training. We adopt cross-entropy loss with padded tokens. A batch size of 30 balances GPU memory utilization with stable gradient estimates. The loss function is the standard cross-entropy between the predicted pointer logits and the target sequence. To prioritize track termination prediction, all loss terms corresponding to the separation \texttt{[TRACK]} token are scaled by a fixed weight factor of 30. All models are trained on a single NVIDIA GeForce RTX 4090 GPU (24 GB VRAM). Training duration varies from 50 to 200 epochs depending on sample size.

At inference, greedy autoregressive decoding is performed under hard physical and structural constraints. Previously selected and padded hits are masked, and consecutive \texttt{[TRACK]} tokens are prohibited. Since the output structure is constrained by the requirement that each valid hit be assigned once and that the number of track groups be fixed, the sequence length is determined by the event-level constraints. Therefore, no explicit learned \texttt{[END]} token is required; decoding terminates once the required output structure is completed. A final constrained post-processing step is applied when necessary to enforce the required number of track boundaries.

\subsection{Evaluation Metrics}

To evaluate our model, we adopt the metrics defined in Ref.~\cite{Qian:2026pbd}, which were originally developed for DC-related tracking. Hit efficiency, denoted as $\epsilon_{\text{hit}}$, represents the fraction of a particle's detectable true hits that are correctly reconstructed and matched to that particle. Hit purity, $p_{\text{hit}}$, measures the fraction of reconstructed hits assigned to a track that are correctly matched to their originating particle. Track efficiency, $\epsilon_{\text{track}}$, is defined as the fraction of detectable true tracks for which a matched reconstructed track exists. A reconstructed track is considered to be a matched track if it satisfies the track-matching criteria: $p_{\text{hit}} > 0.50$, $\epsilon_{\text{hit}} > 0.20$, and $N_{\text{hit}}^{\text{matched}} \geq 6$. Finally, clone or fake rate, $R_{\text{clone+fake}}$, is defined as the total count of clone or fake tracks divided by the total count of detectable truth tracks.

%=============================================================================
\section{Experiments and Results}
%=============================================================================

To systematically evaluate the model, we design our experiments around three tests. First, we assess the model using samples with well-defined track topologies, referred to as Sample 1 to Sample 5, as listed in Table~\ref{tab:datasets}. Second, we examine the performance of model on Sample 6, which contains different track topologies, comparing results with and without prior knowledge of the number of tracks. Finally, we test the model on Sample 7, which includes long-lived particles.

\subsection{Single-Track and Multi-Track Performance}

\begin{figure}[htbp]
    \centering
    \includegraphics[width=0.45\textwidth]{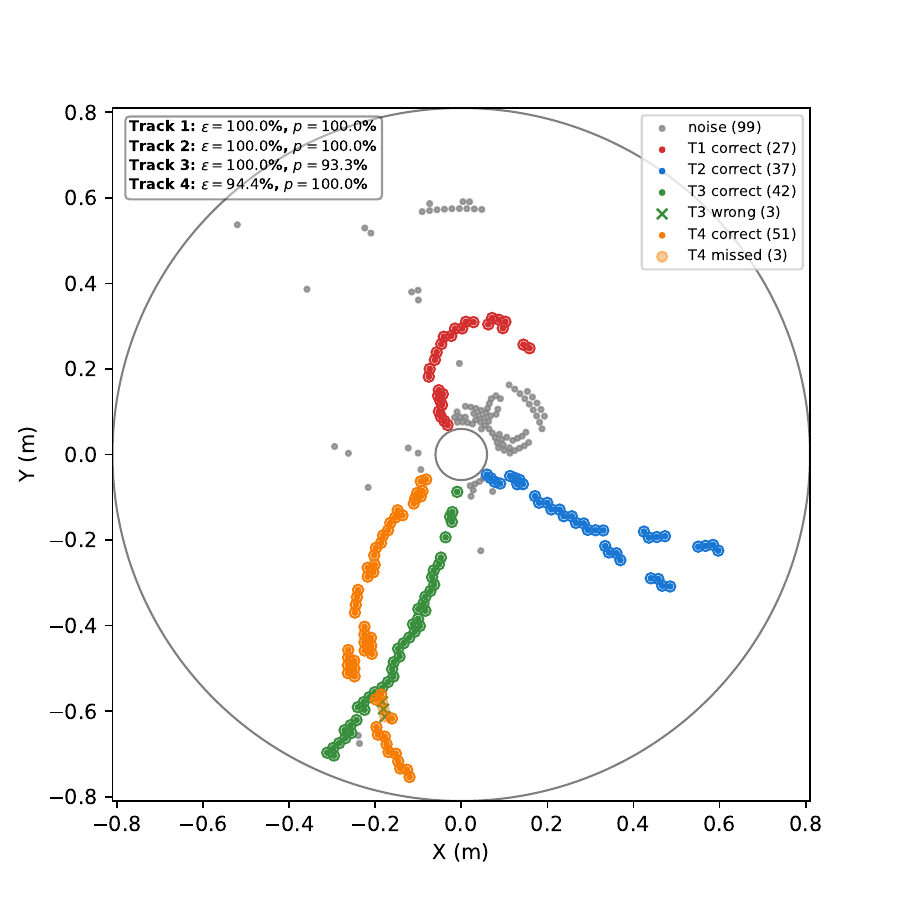}
    \put(-40, 30){\bf (a)}
    \includegraphics[width=0.45\textwidth]{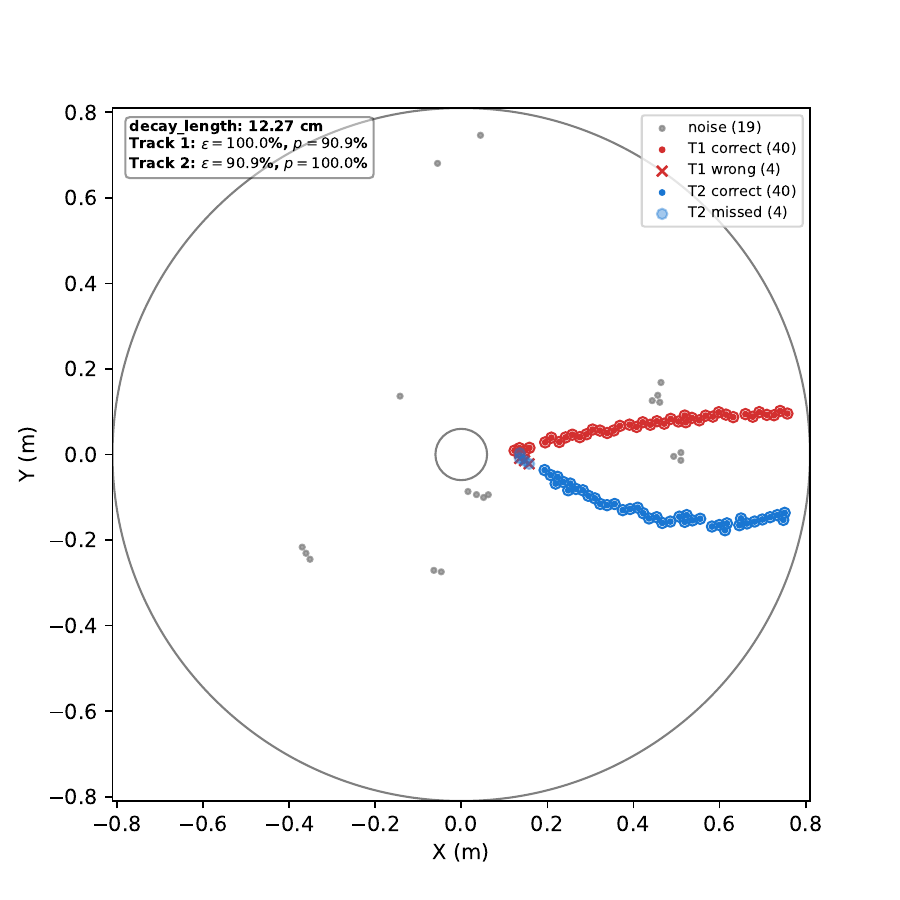}
    \put(-40, 30){\bf (b)}
    \caption{The predicted hit assignments for (a) $\pi^+\pi^-J/\psi(\to e^+e^-)$ event and (b) $K_S^0 \to \pi^+\pi^-$ event. To show all possible results of the model prediction, these two examples include correctly predicted hits, missed hits, and wrong hits, each shown with different marker styles. Noise hits are shown as gray dots.}
    \label{fig:evt}
\end{figure}

We begin with Samples 1 to 4, which provide clean single-track and multi-track events with varying complexity. A separate model is trained on instance on each sample. Figure~\ref{fig:evt} (a) shows the predicted hit assignments for a $\pi^+\pi^-J/\psi(\to e^+e^-)$ event, which has four tracks in the final states. The model can clearly distinguish most of the hits from the background and the signal, but it may mis-classify events near the end of the sequence and at the intersections of tracks. Table~\ref{tab:results_t1} presents both the hit-level and the track-level results for all four samples.

The model achieves a hit efficiency and track efficiency of approximately $98\%$ to $99\%$ across all four samples. In addition, it produces very low clone and fake rates. The results on both single-track and multi-track events surpass the hit-level performance of the traditional baseline finder, and exhibit comparable performance at the track level~\cite{Qian:2026pbd}, as listed in Table~\ref{tab:results_t1}. The result on low-$p_T$ electrons (Sample 5) merits special attention. These tracks have transverse momenta of only 40--160~MeV/$c$, meaning they undergo significant multiple scattering in the detector material and follow strongly curved trajectories. Traditional reconstruction algorithms that rely on straight-line or mildly curved track models struggle in this regime~\cite{Jia:2010zz}. Our model shows good performance here ($\epsilon_{\text{track}} = 98\%$), suggesting that the BERT encoder's global attention mechanism is resilient to non-ideal track shapes: rather than imposing a fixed geometric model, it learns the actual hit correlation patterns present in the data.

\begin{table}[htbp]
    \centering
    \caption{Hit efficiency ($\epsilon_{\text{hit}}$), hit purity ($p_{\text{hit}}$), track efficiency ($\epsilon_{\text{track}}$), and clone or fake rate ($R_{\text{clone+fake}}$) for Samples 1--5 and results from the baseline finder~\cite{Qian:2026pbd}. This results in percent are based on test sets, and uncertainties are statistical.}
    \label{tab:results_t1}
    \begin{tabular}{lcccc}
    \toprule
    Sample & $\epsilon_{\text{hit}}$ & $p_{\text{hit}}$ & $\epsilon_{\text{track}}$ & $R_{\text{clone+fake}}$ \\
    \midrule
S1 ($e^{\pm}$) & $99.76 \pm 0.02$ & $99.60 \pm 0.03$ & $99.73 \pm 0.16$ & $<0.01$ \\
S2 ($\pi^{\pm}$) & $99.73 \pm 0.03$ & $99.52 \pm 0.03$ & $99.73 \pm 0.17$ & $0.10 \pm 0.10$ \\
S3 ($\mu^+\mu^-$) & $98.80 \pm 0.04$ & $98.67 \pm 0.04$ & $98.51 \pm 0.28$ & $0.21 \pm 0.11$ \\
S4 ($\pi^+\pi^-e^+e^-$) & $99.33 \pm 0.02$ & $99.28 \pm 0.02$ & $98.96 \pm 0.17$ & $0.30 \pm 0.09$ \\
S5 ($e^{\pm}$ low $p_T$) & $99.55 \pm 0.01$ & $99.38 \pm 0.01$ & $99.55 \pm 0.07$ & $0.05 \pm 0.02$ \\
    \midrule
Baseline Finder ($\pi^{\pm}$)~\cite{Qian:2026pbd} & $92.24 \pm 0.12$ & $98.58 \pm 0.05$ & $99.71 \pm 0.02$ & $0.08 \pm 0.01$ \\
Baseline Finder ($\pi^+\pi^-$)~\cite{Qian:2026pbd} & $90.87 \pm 0.14$ & $97.93 \pm 0.05$ & $99.63 \pm 0.03$ & $0.11 \pm 0.02$ \\
    \bottomrule
    \end{tabular}
\end{table}

\subsection{Mixed-Sample Performance}

The following test is based on mixed Sample 6, which combines events from Samples 1 to 5, and includes different multiplicities. We also investigated whether providing the true number of tracks as auxiliary information at inference time improves performance. When the track count is specified, the model uses it to determine how many \texttt{[TRACK]} tokens to predict, providing a soft constraint on the output structure. Beside, we also evaluate the model by allowing it to freely predict the number of tracks.

Table~\ref{tab:results_t2} compares the mixed-trained model with and without specifying the number of tracks. The model achieved roughly comparable performance under the two strategies, suggesting that it had learned the underlying information from the data and did not strongly rely on tracking-related prior knowledge. Moreover, the fact that incorporating the prior knowledge of the number of tracks did not further improve performance also suggests that the bottleneck may lie in the model itself and the size of the dataset, rather than in the training strategy.

\begin{table}[htbp]
    \centering
    \caption{Hit efficiency ($\epsilon_{\text{hit}}$), hit purity ($p_{\text{hit}}$), track efficiency ($\epsilon_{\text{track}}$), and clone or fake rate ($R_{\text{clone+fake}}$) for mixed-trained model with and without specifying the number of tracks. This results in percent are based on test sets, and uncertainties are statistical.}
    \label{tab:results_t2}
    \begin{tabular}{lcccc}
    \toprule
    Sample & $\epsilon_{\text{hit}}$ & $p_{\text{hit}}$ & $\epsilon_{\text{track}}$ & $R_{\text{clone+fake}}$ \\
    \midrule
Specified & $99.70 \pm 0.01$ & $99.49 \pm 0.01$ & $99.57 \pm 0.06$ & $0.07 \pm 0.02$ \\
Free & $99.72 \pm 0.01$ & $99.51 \pm 0.01$ & $99.42 \pm 0.07$ & $0.16 \pm 0.03$ \\
    \bottomrule
    \end{tabular}
\end{table}

\subsection{Long-lived Particle Performance}

Finally, we study the performance of model on Sample 7 with $K_S^0 \to \pi^+\pi^-$ events. A case of the predicted hit assignments for a $K_S^0 \to \pi^+\pi^-$ event is illustrated as Fig.~\ref{fig:evt} (b). The $K_S^0$ meson travels a measurable distance (typically several centimeters) before decaying into two charged pions. These pions originate from a displaced secondary vertex, not from the primary $e^+e^-$ interaction point.

\begin{figure}[htbp]
    \centering
    \includegraphics[width=0.60\textwidth]{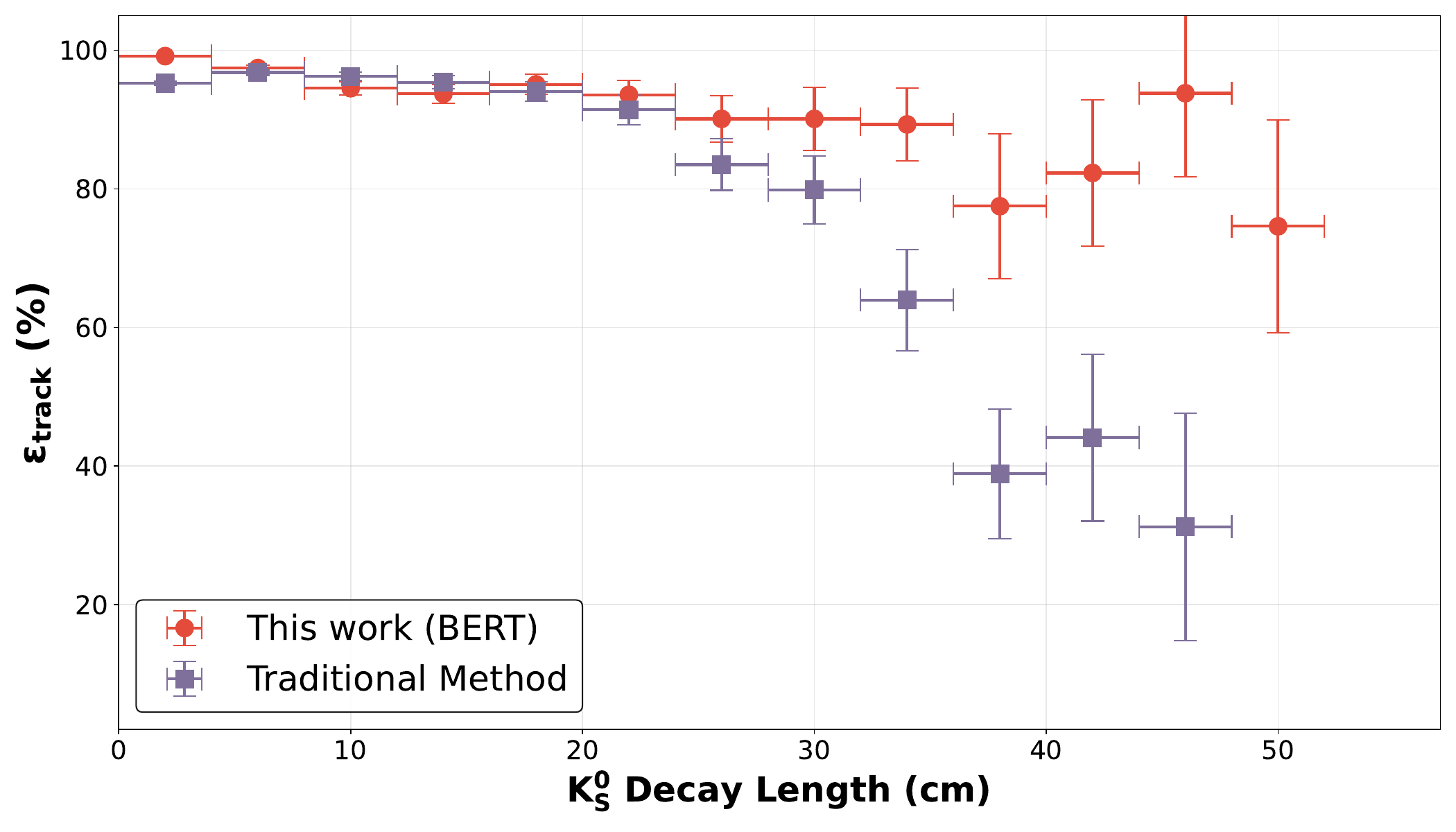}
    \caption{$K_S^0$ event reconstruct efficiency versus $K_S^0$ decay length, compared with the traditional baseline finder.}
    \label{fig:ks}
\end{figure}

The model performs well on the entire Sample 7, achieving a hit efficiency of $94.84 \pm 0.03\%$, a hit purity of $95.51 \pm 0.03\%$, a track efficiency of $97.62 \pm 0.30\%$, and a clone or fake rate of $1.33 \pm 0.10\%$. To further evaluate the performance with lengths between the interaction point and different displaced secondary vertices, Figure~\ref{fig:ks} shows a comparison between our model and the traditional baseline finder in terms of $K_S^0$ event efficiency as a function of the $K_S^0$ decay length in X-Y plane. The track efficiency in the traditional baseline finder is estimated as the ratio of reconstructed tracks. Our model performs comparably to conventional methods at small decay lengths, but its advantage becomes increasingly pronounced at larger decay lengths. This is because conventional methods typically use a global tracking strategy that requires the track to start from the $e^+e^-$ interaction point to better constrain the track parameters. In contrast, the encoder-decoder architecture employs global self-attention instead of a local seeding strategy and does not have an inherent bias toward the primary vertex.

%=============================================================================
\section{Conclusion and Outlook}
%=============================================================================

In summary, we present a BERT-based Transformer model for track reconstruction in drift chambers and evaluate it on the DCTracks open dataset, which contains six event samples with different track multiplicities and kinematic configurations. The model achieves about 98\% hit and track efficiencies while maintaining very low clone and fake rates. Compared with traditional vertex-seeded algorithms, it also improves the reconstruction of displaced vertices. These results demonstrate the effectiveness of the proposed approach for track reconstruction in drift chambers.

This work explores the application of deep learning, in particular Transformer-based sequence models, to track finding in high energy physics. The encoder-decoder architecture is modular, with the encoder providing contextualized hit representations that can serve as a common input for downstream tasks. For example, additional heads could be introduced for particle classification and initial track-parameter prediction to support subsequent track fitting. The model also has some architectural limitations. Since we fine-tune the original BERT~\cite{Devlin:2018mgb}, the input sequence is limited by the maximum BERT sequence length, requiring $N_{\rm noise}+N_{\rm hits}<512-N_{\rm \texttt{[TRACK]}}$. Increasing the input length or adopting a smaller custom Transformer architecture as the backbone could allow the model to handle events with more tracks and more complex backgrounds. Although denoising and track finding are combined in a single end-to-end framework, an additional denoising step before tracking could also be explored to further improve the reconstruction performance. The choice of input representation may also affect the results; for instance, incorporating higher-dimensional representations from GNNs may provide additional information for the Transformer-based model. Furthermore, future studies could relax the constraint on the number of output hits and introduce a learned \texttt{[END]} token to investigate whether the model can learn the termination condition autonomously.

Looking ahead, the proposed approach can be extended to other $\tau$-charm and flavor-factory experiments, such as Belle~II~\cite{Belle-II:2010dht} and STCF~\cite{Achasov:2023gey}, which also feature low background levels and require high-precision measurements. The results obtained with the Transformer-based sequence-to-sequence approach indicate its potential for track reconstruction in high energy physics and provide a complementary strategy to existing methods such as GNN-based approaches. More generally, combining pretrained foundation-model architectures with task-specific components may provide a flexible framework for adapting modern deep-learning models to detector reconstruction problems.

%=============================================================================
\section*{Acknowledgments}
We thank the IHEP Computing Center for providing computing resources. We also thank Shengsen Sun for valuable comments on the draft. This work is supported the National Natural Science Foundation of China (NSFC) under Contracts Nos.~12575207. This work is supported by the Strategic Priority Research Program of the Chinese Academy of Sciences under Grant XDA0480600.

%=============================================================================

\end{document}